\documentclass[twocolumn,prl,superscriptaddress,showpacs]{revtex4}

\usepackage{amsmath,amssymb,graphicx,psfrag,subfloat,times}

\renewcommand\epsilon{\varepsilon}
\renewcommand\phi{\varphi}
\renewcommand\vec[1]{\boldsymbol{\mathrm{#1}}}
\newcommand\diff{\mathrm{d}}

\newcommand\eq[1]{Eq.~\eqref{eq:#1}}
\newcommand\fig[1]{Fig.~\ref{fig:#1}}

\begin{document}

\title{Localization Transition of the 3D Lorentz Model and Continuum Percolation}
\author{Felix H{\"o}f\/ling}
\affiliation{Hahn-Meitner-Institut Berlin, Abteilung Theorie, Glienicker Stra{\ss}e 100, D-14109 Berlin, Germany}

\author{Thomas Franosch}
\affiliation{Hahn-Meitner-Institut Berlin, Abteilung Theorie, Glienicker Stra{\ss}e 100, D-14109 Berlin, Germany}
\affiliation{Arnold Sommerfeld Center and CeNS, Department of Physics, Ludwig-Maximilians-Universit{\"a}t M{\"u}nchen, Theresienstra{\ss}e 37, D-80333 M{\"u}nchen, Germany}

\author{Erwin Frey}
\affiliation{Arnold Sommerfeld Center and CeNS, Department of Physics, Ludwig-Maximilians-Universit{\"a}t M{\"u}nchen, Theresienstra{\ss}e 37, D-80333 M{\"u}nchen, Germany}

\begin{abstract}
The localization transition and the critical properties of the
Lorentz model in three dimensions are investigated by computer
simulations. We give a coherent and quantitative explanation of
the dynamics in terms of continuum percolation theory and obtain
an excellent matching of  the critical density and exponents.
Within a dynamic scaling Ansatz incorporating two divergent length
scales   we achieve data collapse for the mean-square
displacements  and identify the leading corrections to scaling. We
provide  evidence for a divergent non-Gaussian parameter close to
the transition.
\end{abstract}


%{\small accepted for publication in Phys. Rev. Lett.}

\maketitle

Transport in heterogeneous and
disordered media has important applications in many fields of
science including composite materials, rheology, polymer and
colloidal science, and biophysics. Recently, dynamic
heterogeneities and growing cooperative length scales in
structural glasses  have attracted considerable interest
\cite{Bertin:2005,Berthier:2005}. The physics of gelation, in
particular of colloidal particles with short range attraction
\cite{Campbell:2005,Manley:2005,Zaccarelli:2005,Ruzicka:2004}, is
often accompanied by the presence of a fractal cluster
generating sub-diffusive dynamics. It is of fundamental interest to
demonstrate the relevance of such heterogeneous environments on
slow anomalous transport.

The minimal model for  transport of particles through a random
medium of fixed obstacles, is known as Lorentz model, and already
incorporates the generic ingredients for slow anomalous transport.
Earlier, the Lorentz model has played a significant role
 as a testing ground for elaborate  kinetic
theories, shortly after the discovery of long-time tails in
auto-correlation functions for simple liquids in the late
1960s~\cite{Alder:1970}, since the non-analytic dependence of
transport coefficients on frequency, wavenumber, and density
predicted for simple
liquids~\cite{Dorfman:1970,Ernst:1971b,Kawasaki:1971,Bedaux:1973,Tokuyama:1978}
has a close analog in the Lorentz model
\cite{Weijland:1968,Ernst:1971a}.

\begin{figure}[b]
\includegraphics[width=.85\linewidth]{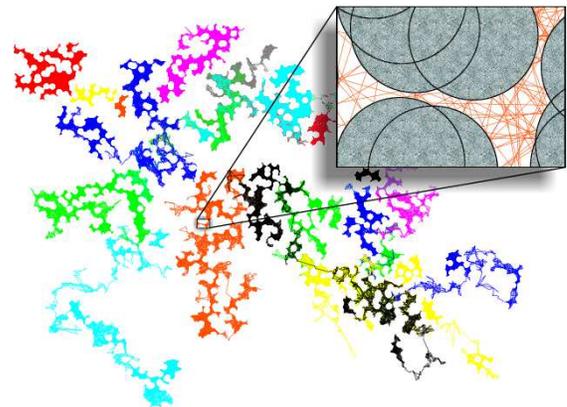}
\caption{(color) Typical particle trajectories in a 2D Lorentz
model slightly below $n^*_c$ over a few thousand collisions each.
Colors encode  different initial conditions; obstacles have been
omitted for clarity. Most trajectories being in the percolating
void space have some overlap; a few trajectories are confined to
finite clusters. Blow-up: a particle squeezes through narrow gaps
formed by the obstacles.} \label{fig:trajectory}
\end{figure}

The simplest variant of the Lorentz model consists of a
structureless test particle moving according to Newton's laws in a
$d$-dimensional array of identical obstacles. The latter are
distributed randomly and independently in space and interact with
the test particle via a hard-sphere repulsion. Consequently, the
test particle explores a disordered environment of possibly
overlapping regions of excluded volume; see \fig{trajectory}. Due
to the hard-core repulsion, the magnitude of the particle
velocity,  $v=|\vec v|$, is conserved. Then, the only control
parameter is the dimensionless obstacle density, $n^*:=n\sigma^d$,
where $\sigma$ denotes the radius of the hard-core potential. At
high densities, the model exhibits a localization transition,
i.\,e., above a critical density, the particle is always trapped
by the obstacles.

Significant insight into the dynamic properties of the Lorentz
model has been achieved by a low-density expansion for the
diffusion coefficient by \textcite{Weijland:1968} rigorously
demonstrating the non-analytic dependence on $n^*$. As expected,
for low densities the theoretical results compare well with
Molecular Dynamics (MD) simulations~\cite{Bruin:1974}. Elaborate
self-consistent kinetic theories \cite{Goetze:1981,Masters:1982}
have allowed going much beyond such perturbative approaches. They
give a mathematically consistent description of the localization
transition, which allows to calculate the critical density within
a $20\%$ accuracy and  extend  the regime of quantitative
agreement  to intermediate densities. In addition, they have
provided a microscopic approach towards anomalous transport and
mean-field-like scaling behavior \cite{Goetze:1981}. %\cite{Goetze:1981a,Goetze:1981b}.

A different line of approach focusing on the localization
transition starts from the fractal nature of the void space
between the overlapping spheres in the Lorentz model and considers
it as a continuum percolation problem
\cite{Kertesz:1981,Elam:1984,Halperin:1985,Machta:1985,Stenull:2001},
which in this context has also been termed ``Swiss cheese''
model~\cite{Halperin:1985}. These authors conjectured that the
transport properties close to the percolation threshold can be
obtained by analyzing an equivalent random resistor network. The
equivalence, however, has been shown only for geometric properties
close to the percolation point \cite{Kerstein:1983}. As a
peculiarity of continuum percolation, differences to lattice
percolation may arise due to power law tails in the probability
distribution of the conductances (``narrow gaps''). Such random
resistor networks have been investigated extensively by means of
Monte-Carlo simulations \cite{Derrida:1984,Gingold:1990} and
renormalization group techniques \cite{Harris:1984,Lubensky:1986},
providing reliable numeric and analytic results for the critical
behavior \cite{Havlin:1987}.

In this Letter, we present a direct numerical analysis of the
dynamic properties of the Lorentz model without resorting to
random resistor networks. By means of extensive MD simulations, we
obtain a quantitative description of the dynamic properties over
the full density range, in particular, focusing on both sides of
the critical region. This allows for a quantitative test of the
conjectured mappings to continuum percolation theory. Furthermore,
we explore the range of validity of the dynamic scaling hypothesis
for the Lorentz model \cite{Kertesz:1983}. The probability
distribution of particle displacements, i.\,e., the van Hove
self-correlation function, $G(\vec
r,t):=\big\langle\delta\boldsymbol(\vec r - \Delta \vec
R(t)\boldsymbol)\big\rangle$, and its second moment, the
mean-square displacement (MSD), $\delta
r^2(t):=\left\langle|\Delta\vec R(t)|^2\right\rangle$, are the
appropriate quantities  for this purpose; $\Delta\vec R(t)=\vec
R(t)-\vec R(0)$ denotes the displacement of the test particle at
time $t$. \subfiguresbegin
\begin{figure}
\psfrag{Time}[c][c]{Time $t/v^{-1}\sigma$}
\psfrag{Mean-squared displacement}[c][c]{$\delta r^2(t)/\sigma^2$}
\psfrag{Density}[][]{\footnotesize Density $n^*$~}
\psfrag{L\_box\^2}[bl][bl]{$L_\text{box}^2$}
\psfrag{critical}[ct][ct]{$\delta r^2(t)\sim t^{2/z}$}
\includegraphics[width=\linewidth]{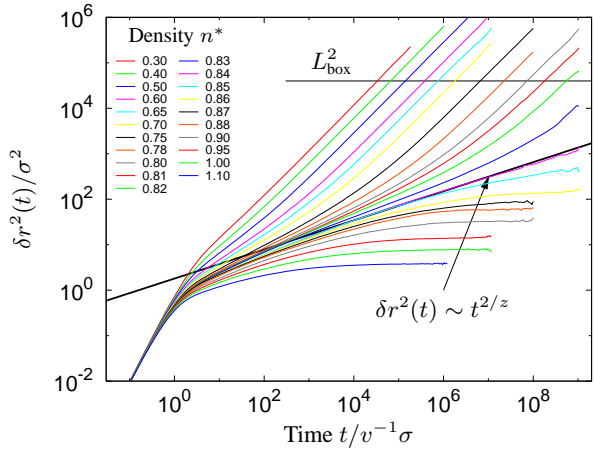}
\caption{(color) MSD $\delta r^2(t)$ for various obstacle densities $n^*$
varying from $0.30$  (top) to $1.10$ (bottom). The thick black
line represents a power law, $\delta r^2(t)\sim t^{2/z}$ with $z =
6.25$.}
\label{fig:msd3d}
\end{figure}

Over a wide range of obstacle densities, we have simulated several
hundred trajectories in three dimensions, %up to more than a billion collisions each,
employing an event-oriented MD algorithm. For each of $N_r$
different realizations of the obstacle disorder, a set of $N_t$
trajectories with different initial conditions is simulated. Below
the critical density, we have chosen $N_r\ge 25$ and $N_t\ge 4$.
At very high densities, where the phase space is highly
decomposed, these values have been increased up to $N_r\times
N_t=600$. In order to minimize finite-size effects, the size of
the simulation box, $L_\text{box}$, has been chosen significantly
larger than the correlation length $\xi$,
$L_\text{box}=200\sigma\gg \xi$ \footnote{This relation may be
violated for $|\epsilon|<0.01$. We checked that our findings are
not affected by finite-size effects. A detailed analysis will be
presented elsewhere.}.

\begin{figure}
\psfrag{scaling variable x=t/te}[t][t]{Scaling variable $\hat t\sim t\ell^{-z}$}
\psfrag{f(x)}[b][b]{$t^{-2/z}\,\delta r^2(t)\,/\,(1+C t^{-y})$}
\psfrag{n > nc}[l][l]{\footnotesize $n^*>n^*_c$}
\psfrag{n < nc}[l][l]{\footnotesize $n^*<n^*_c$}
\psfrag{C=0}[l][l]{\footnotesize $C=0$}
\psfrag{C=-0.8}[l][l]{\footnotesize $C=-0.8$}
\includegraphics[width=\linewidth]{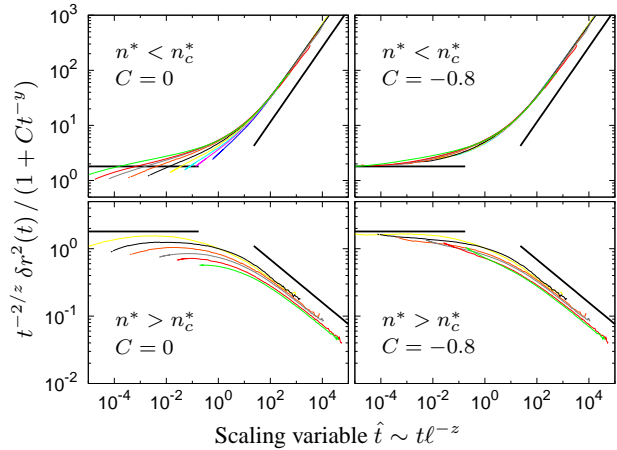}
\caption{(color) Scaling functions $\delta \hat r^2_\pm(\hat t)$ for the
MSD. Right panels include corrections to scaling at leading order.
Units are chosen such that $v=\sigma=1$; color key as in
\fig{msd3d}.} \label{fig:msd_scaling}
\end{figure}
\subfiguresend

The results for the MSD cover a non-trivial time window of more
than seven decades for densities close to the transition, see
\fig{msd3d}. At low densities, one observes only a trivial
cross-over from ballistic  motion, $\delta r^2(t)=v^2 t^2$, to
diffusion, $\delta r^2(t)\sim t$, near the mean collision time
$\tau=1/\pi n v \sigma^2$ as expected from  Boltzmann theory.
 With increasing density, an intermediate time
window opens where motion becomes sub-diffusive, $\delta r^2(t)\sim
t^{2/z}$ with $z>2$. This time window extends to larger and larger
times upon approaching a certain critical density $n^*_c$. For the
density $n^*=0.84$, the sub-diffusive behavior is obeyed over more
than five decades and is compatible with a value of
$z\approx6.25$. The power law, $\delta r^2(t)\sim t^{2/z}$,
indicated in \fig{msd3d},  discriminates nicely  trajectories
above and below $n^*_c$. One also observes a density-dependent
length scale $\ell$ characterizing the end of the sub-diffusive
regime by $\delta r^2(t)\simeq \ell^2$; upon approaching $n^*_c$
this cross-over length scale $\ell$ is found to diverge. For long
times, the dynamics eventually becomes either diffusive or
localized for densities below or above $n^*_c$, respectively.

\begin{figure}
\psfrag{Diffusion coefficient D/D\_0}[c][c]{Diffusion coefficient
$D/D_0$} \psfrag{Density n}[c][c]{Density $n^*$}
\psfrag{|Epsilon|}[c][t]{\footnotesize $|\epsilon|=|n-n_c|/n_c$}
\psfrag{D}[c][c]{\footnotesize $D/v \sigma$}
\psfrag{l}[t][t]{$\ell/\sigma$}
\psfrag{D\~|e|\^mu}[c][c]{\footnotesize $\begin{array}{c}D\sim
|\epsilon|^\mu \\ \mu= 2.88\end{array}$}
\psfrag{l\~|e|\^-nu+beta/2}[cb][cb]{$\ell \sim
|\epsilon|^{-\nu+\beta/2}$} \psfrag{this work}{\footnotesize this
work} \psfrag{Bruin}{\footnotesize \textcite{Bruin:1974}}
\includegraphics[width=\linewidth]{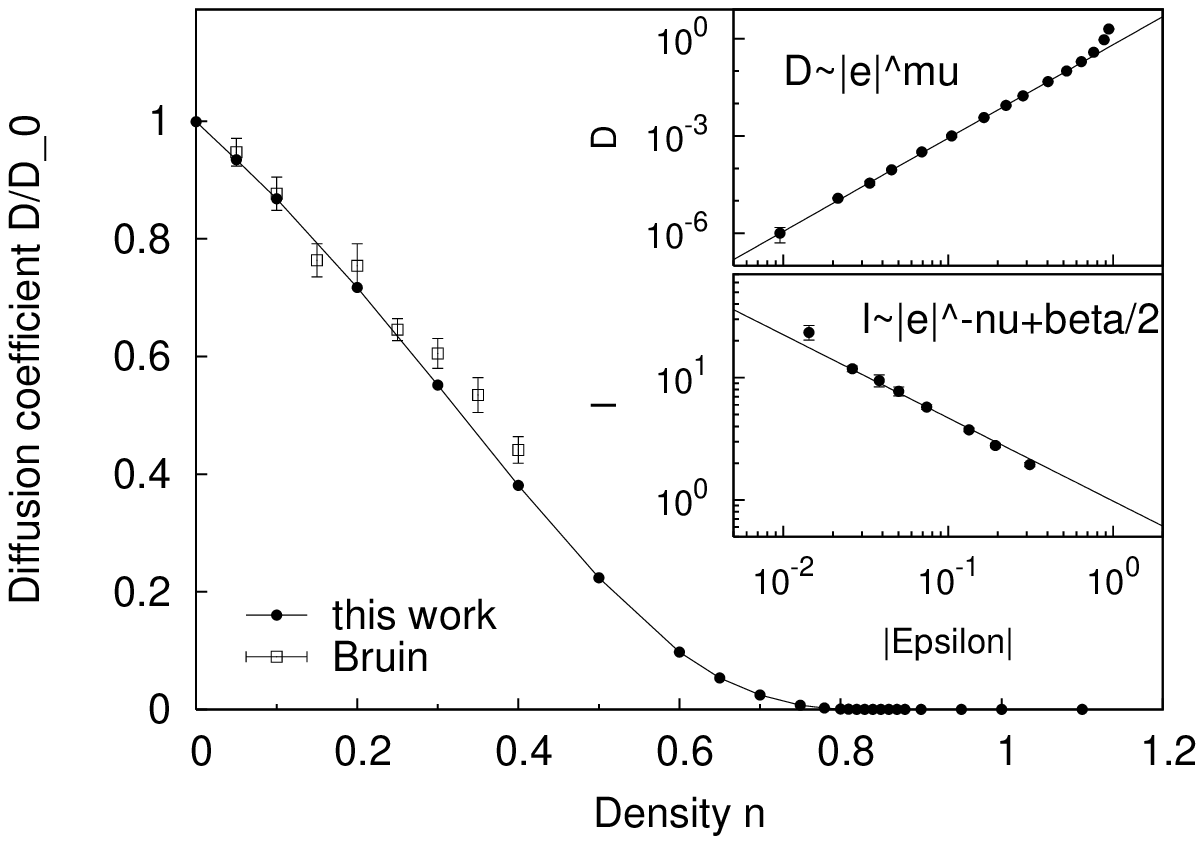}
\caption{Suppression of the diffusion coefficient $D/D_0$ with
increasing density $n^*$. Top inset: Power-law behavior of $D$
close to $n^*_c$. Bottom inset: Divergence of the localization
length $\ell$ upon approaching $n^*_c$ with exponent $\nu-\beta/2=
0.68$.} \label{fig:diffusion_coefficient3d}
\end{figure}

The diffusion coefficient $D$ has been extracted from the
long-time limit of $\delta r^2(t)/6t$; in
\fig{diffusion_coefficient3d}, $D$ is shown in units of the
Boltzmann result, $D_0=\tau v^2/3$. With increasing density, $D$
is more and more suppressed until it vanishes at $n^*_c$ as a
power law, $D\sim|\epsilon|^\mu$, where
$\epsilon:=(n^*-n^*_c)/n^*_c$ defines the separation parameter.
Anticipating the exponent $\mu$ from percolation theory, a fit to
our data yields the critical density, $n^*_c=0.839(4)$
\footnote{This value for $n^*_c$ corresponds to a critical volume
fraction for the obstacles, $\phi_c=1-\exp(-\frac{4\pi}{3}
n^*_c)=0.9702(5)$.}, and the power law behavior is confirmed over
five decades in $D$. Above the critical density, the long-time
limit of the MSD is compatible with a power law over more than one
decade, $\ell\sim\epsilon^{-\nu+\beta/2}$, where
$\nu-\beta/2\approx 0.68$ (bottom inset in
\fig{diffusion_coefficient3d}). Our finding of $n^*_c$ coincides
with the percolation point of the void space
\cite{Kertesz:1981,Elam:1984,Rintoul:2000}. This provides clear
evidence for the intimate connection between continuum percolation
and the Lorentz model, i.\,e., diffusion is not blocked as long as
there is an infinite path through the medium---a purely geometric
reason.

Considering the underlying continuum percolation problem, a geometric transition occurs at $n^*_c$,
above which the void space falls completely apart into finite
clusters. Just below this density, the volume fraction $P$ of the
percolating void space (infinite cluster) vanishes as a power law,
$P\sim|\epsilon|^\beta$. There are two divergent length scales
characterizing the structure of the percolation network: the
linear dimension of the largest finite clusters, $\xi\sim|\epsilon|^{-\nu}$,
 and the mean cluster radius (radius of gyration),
$\ell\sim|\epsilon|^{-\nu+\beta/2}$ \cite{Stauffer:Percolation}.
The geometric exponents $\beta$ and $\nu$ are believed to be the
same for lattice and continuum percolation~\cite{Elam:1984}. Our
results in \fig{diffusion_coefficient3d} clearly identify the
geometric mean cluster radius $\ell$ with the localization length
of the MSD as anticipated by our choice of notation.

In continuum percolation, transport of a particle is limited by
narrow gaps in the void space. It was argued that this feature of
the dynamics is captured by an associated random resistor network
with a  distribution $\rho(W)$ of weak
conductances $W$ exhibiting  a power-law tail, $\rho(W)\sim
W^{-\alpha}$, $\alpha<1$ for small $W$
\cite{Halperin:1985,Machta:1985}. Depending on the value of
$\alpha$, the suppression of diffusion, $D\sim|\epsilon|^\mu$,
may be dominated by this tail, and dynamic exponents become
different from lattice percolation, $\mu > \mu^{\text{lat}}$. In this case,
 the hyperscaling relation, $\mu = (d-2)\nu + 1/(1-\alpha)$, holds
\cite{Straley:1982,Stenull:2001}.
%The parameter $\alpha$ has to be
%inferred from the geometrical properties of the Lorentz model.
There is a discrepancy in the literature about the value of
$\alpha$ in the Lorentz model
\cite{Halperin:1985,Machta:1985,Havlin:1987}. Only the result of
\textcite{Machta:1985}, $\alpha=(d-2)/(d-1)$, is consistent with
our data. In $d=3$, it implies   $\mu=\nu+2\approx 2.88$, and
therefore, $\mu>\mu^\text{lat} \approx 2.0$ \footnote{All exponents
are calculated consistently based on the values $\beta=0.41$,
$\nu=0.88$, and $\mu^\text{lat}=2.0$ \cite{Stauffer:Percolation}.}. By means of a scaling relation
\cite{Stauffer:Percolation}, $z=(2\nu-\beta+\mu)/(\nu-\beta/2)$,
one finds the dynamic exponent, $z \approx 6.25$,  describing
anomalous transport at critical density, $\delta r^2(t)\sim
t^{2/z}$. Note that this dynamic exponent is not independent but
entirely determined by the geometric properties of the random
environment.

In conclusion, the values obtained from the simulated MSD for the
critical density $n^*_c$, the dynamic exponent  $z$ as well as
the  exponents for the diffusion coefficient $\mu$ and the
localization length $\nu-\beta/2$ agree with the predicted values
for continuum percolation. Within the statistical accuracy,
no deviations can be inferred.

The quality of our data allows to go beyond determining critical
exponents and to give a full analysis of the dynamic scaling
properties. It has been argued by \textcite{Kertesz:1983} that the
van Hove correlation function obeys scaling. Rewriting their
Ansatz in a more transparent way yields,
\begin{equation}
G(\vec r,t;\epsilon)=\xi^{-\beta/\nu-d}{\cal G}_\pm(\vec r/\xi,t\ell^{-z}),
\label{eq:vanHove_scaling}
\end{equation}
where  ${\cal G}_\pm$ are master functions  above ($+$) and below
($-$) the critical density.  This Ansatz clearly reflects the
r{\^o}le of the two length scales: the correlation length $\xi$
rescales geometry whereas the cross-over length scale $\ell$
rescales time. The scaling form of the MSD is easily inferred from
$\delta r^2(t; \epsilon)= \int\!\diff^d\vec r\,r^2 G(\vec r,t;
\epsilon)$ as, $\delta r^2(t;\epsilon)=t^{2/z}\delta \hat
r^2_\pm(\hat t),$ where $\hat t\sim t\ell^{-z}$. Plotting
$t^{-2/z}\delta r^2(t;\epsilon)$ versus $\hat t$ for various
densities (left panels of \fig{msd_scaling}), the  data collapse
nicely in the diffusive and localized  regimes ($\hat t\gg 1$) and
converge rapidly to the corresponding large-$\hat t$ asymptotes,
$\delta \hat r^2_-(\hat t)\sim \hat t^{1-2/z}$ and $\delta \hat
r^2_+(\hat t)\sim \hat t^{-2/z}$.  Convergence to the critical
asymptote, $\delta \hat r^2_\pm(\hat t)\sim\mathit{const}$, for
$\hat{t} \ll 1$ becomes increasingly better as the critical point
is approached.

Deviations from scaling can be rationalized  by considering the
again universal corrections to scaling. Extending the Ansatz,
\eq{vanHove_scaling}, by an irrelevant parameter $u$ leads to
$\delta r^2(t;\epsilon,u)=t^{2/z}\Phi_\pm(t\ell^{-z},u t^{-y})$,
where $y$ is a universal exponent. Since $\Phi_\pm$ is assumed to
be analytic for small arguments, one obtains the leading order
correction upon expanding $\Phi_\pm$ to first order in $u$,
\begin{equation}\label{eq:corrections}
\delta r^2(t;\epsilon)=t^{2/z}\delta \hat r^2_\pm(\hat
t)\boldsymbol(1+t^{-y}\Delta_\pm(\hat t)\boldsymbol) \, ,
\end{equation}
introducing some analytic functions $\Delta_\pm(\hat{t})$.
Specializing \eq{corrections} to the critical density, i.\,e.,
$\hat{t}=0$, yields $\delta r^2(t;\epsilon=0)\propto t^{2/z} (1+ C
t^{-y} )$, with a single amplitude $C = \Delta_{\pm}(\hat{t} =
0)$; it also identifies $y$ as the leading non-analytic correction
exponent at criticality. Our data for $n^*_c = 0.84$ are
compatible with values for $y$ between $0.15$ and $0.4$.
For the following, we found the choice $y=0.34$ and $C=-0.8$
reasonable, the value for $y$ is supported by theoretical arguments to be
presented elsewhere.

Inspection of \fig{msd_scaling} reveals that corrections to
scaling are less relevant for long times, $\hat{t}\gg 1$, whereas
significant deviations are visible in the critical regime,
$\hat{t} \ll 1$. This observation is consistent with the scaling
behavior of the diffusion coefficient and the localization length,
see \fig{diffusion_coefficient3d}. These findings also suggest approximating
the corrections by its value at $\hat{t}=0$, i.\,e.,
substituting $\Delta_\pm(\hat{t}) = C$  in \eq{corrections} for all times,
$ \delta r^2(t;\epsilon)=t^{2/z}\delta \hat
r^2_\pm(\hat t)\,\big(1+C t^{-y} \big)$. With $y$ and $C$ already
inferred from the data close to criticality, the correction terms
should apply for all densities. Indeed, including this
leading-order correction improves the data collapse substantially
(\fig{msd_scaling}, right panels).

\begin{figure}
\psfrag{Time}[t][t]{Time $t/v^{-1}\sigma$}
\psfrag{Mean-quartic displacement}[b][b]{$\delta r^4(t)/\sigma^4$}
\psfrag{Time2}[t][t]{\footnotesize $t/v^{-1}\sigma$}
\psfrag{NGP}[l][l]{$\alpha_2(t)$}
\psfrag{Density}[c][c]{\footnotesize Density $n^*$}
\psfrag{t\^4/z'}[l][l]{\small $t^{4/\tilde z}$}
\psfrag{t\^4/z}[l][l]{\small $t^{4/z}$}
\psfrag{n=0.70}[rb][rb]{\scriptsize $n^*=0.70$}
\psfrag{n=1.00}[rb][rb]{\scriptsize $n^*=1.00$}
\includegraphics[width=\linewidth]{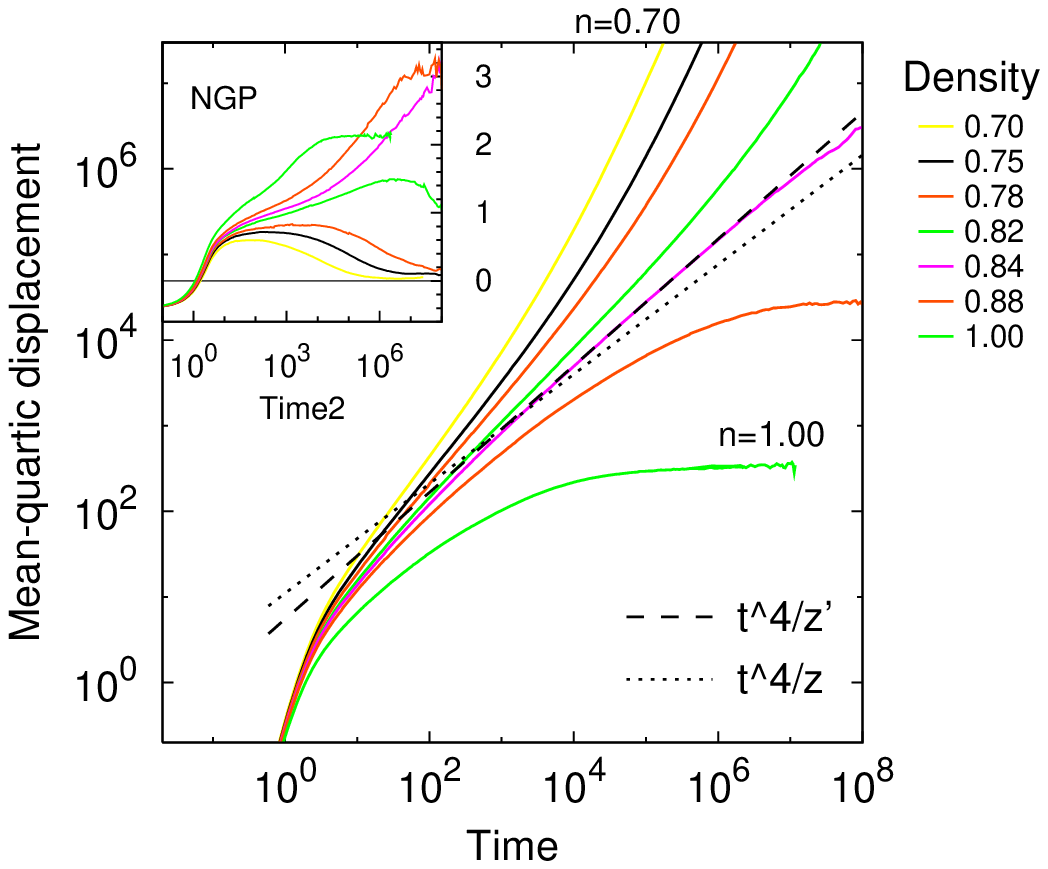}
\caption{(color online) MQD $\delta r^4(t)$ for densities above
and below $n^*_c$. The dashed and dotted lines compare the two
exponents $\tilde z$ and $z$. Inset: NGP $\alpha_2(t)$ for the
same densities, its long-time limit increases as $n^*_c$ is
approached.} \label{fig:mqd-ngp3d}
\end{figure}

The presence of two different length scales, $\ell$ and $\xi$, in
the scaling hypothesis, \eq{vanHove_scaling}, is not manifested in
the MSD; it will, however, affect the higher moments of the
probability distribution, e.\,g., the mean-quartic displacement
(MQD), $\delta r^4(t; \epsilon)= \int\!\diff^d\vec r\,r^4 G(\vec
r,t; \epsilon)$. Above $n^*_c$, it is easily inferred that the
long-time limit of the MQD scales as, $\delta
r^4(t\to\infty)\sim\xi^2\ell^2$. At the critical density, we
obtain the long-time asymptote, $\delta r^4(t)\sim t^{4/\tilde
z}$, with the exponent $\tilde
z:=(2\nu-\beta+\mu)/(\nu-\beta/4)\approx 5.45$ different from $z$.
We have evaluated the MQD by our simulation and find agreement
with the prediction of continuum percolation at a similar level of
significance as for the MSD, see \fig{mqd-ngp3d}. In particular,
for the density $n^* =0.84$ the MQD follows a power-law with the
predicted exponent $\tilde z$ for a time window of more than four
decades.

A more sensitive quantity is the (first)
non-Gaussian parameter (NGP), $\alpha_2(t):=\frac{3}{5} \delta
r^4(t) \boldsymbol(\delta r^2(t)\boldsymbol)^{-2}-1,$ quantifying
deviations from a Gaussian distribution \cite{BoonYip:1980}. At
criticality, it diverges as $\alpha_2(t)\sim t^{4/\tilde
z-4/z}\approx t^{0.097}$; direct observation of this very small
exponent is expected to be a considerably difficult task. The
long-time limits of $\alpha_2(t)$ diverge upon approaching $n^*_c$
from either above or below as
$\alpha_2(t\to\infty)\sim|\epsilon|^{-\beta}$. In particular, the
NGP does not vanish in the diffusive regime close to the
transition due to the presence of localized particles even below
$n^*_c$. Although there are significant statistical errors in the
data for the NGP, \fig{mqd-ngp3d} (inset) provides evidence for a significant
increase of $\alpha_2(t)$ as density approaches $n^*_c$ from
either side. The properties of the NGP demonstrate that the
presence of two divergent length scales is crucial for the
understanding of the dynamics close to the localization
transition.

\begin{acknowledgements}
We are indebted to W. G{\"o}tze for valuable discussions and critical comments
 as well as W. Kob and A. Moreno for helpful correspondence.
\end{acknowledgements}

\bibliography{dissertation}

\end{document}